\documentstyle[prl,aps,multicol]{revtex}
\input epsf
\begin{document}
\draft

\title
      {
        Molecular dynamics simulation study of the high frequency sound
        waves \\ in the fragile glass former ortho-terphenyl.
      }
\author{
        S.~Mossa$^{1,2}$,
        G.~Monaco$^{3}$,
        G.~Ruocco$^{2}$,
        M.~Sampoli$^{4}$,
        and F.~Sette$^{3}$
                }
\address{
         $^1$
         Center for Polymer studies and Department of Physics, Boston
         University, Boston, Massachusetts 02215.\\
         $^2$
         Dipartimento di Fisica and INFM, Universit\`a di Roma "La Sapienza"
         P.zza Aldo Moro 2, Roma, I-00185, Italy \\
         $^3$
         European Synchrotron Radiation Facility,
         BP220, Grenoble Cedex, F-38043, France \\
         $^4$
         Dipartimento di Energetica and INFM, Universit\`a di Firenze,
         Via Santa Marta 3 , Firenze, I-50139, Italy \\
        }

\date{\today}

\maketitle
\begin{abstract}

Using a realistic flexible molecule model of the fragile glass
former orthoterphenyl, we calculate via molecular dynamics
simulation the collective dynamic structure factor $S(Q,\omega)$,
recently measured in this system by Inelastic X-ray Scattering.
The comparison of the simulated and measured dynamic structure
factor, and the study of the $S(Q,\omega)$ in an extended
momentum ($Q$), frequency ($\omega$) and temperature range
allows: {\it i)} to conclude that the utilized molecular model
gives rise to $S(Q,\omega)$ in agreement with the experimental
data, for those thermodynamic states and $Q$ values where the
latter are available; {\it ii)} to confirm the existence of a
slope discontinuity on the $T$ dependence of the sound velocity
that, at finite $Q$'s, takes place at a temperature $T_x$ higher
than the calorimetric glass transition temperature $T_g$; {\it
iii)} to find that the values of $T_x$ is $Q$-dependent and that
its $Q\rightarrow$0 limit is consistent with $T_g$. The latter
finding is interpreted within the framework of the current
description of the dynamics of supercooled liquids in terms of
exploration of the potential energy landscape.
\end{abstract}
\pacs{PACS numbers: 64.70.Pf, 71.15.Pd, 61.25.Em, 61.20.-p}

\begin{multicols}{2}

\section{INTRODUCTION}
\label{introduction}

The understanding of the dynamics of the supercooled liquids
\cite{angell_1,angell_2} and of its relation with the glass
transition has received great attention in the last few years.
Different details of the dynamics of supercooled liquids and
glasses seem to be on the way to be settled, while other are still
obscure. As an example, considering the dynamics of the density
fluctuations which are the main issue for this paper, on one side
the connection between the glass transition phenomenology and the
long time dynamics ({\it structural rearrangement}) has been
almost clarified, on the other side the effect of the structural
arrest on the high frequency collective vibrational motion ({\it
sound waves}) is much less clear.

From the experimental point of view the study of the time
behavior of the density fluctuations is made via the determination
of the dynamic structure factor $S(Q,\omega)$, i.~e. the power
spectrum of the $Q$ component of the number density $\rho_Q(t)$.
In the hydrodynamic regime this quantity is measurable by laser
light scattering, a well established technique ($Q\approx$0.03
nm$^{-1}$ corresponding to frequency in the GHz range). Thanks to
this technique much is known about the collective dynamics in the
nanosecond time scale. The limit of high frequencies (picosecond
time scales) is traditionally the realm of the inelastic neutron
scattering technique. This technique, however, suffers of strong
kinematics limitations, that prevent the possibility to
investigate the mesoscopic $Q$ region ($Q\approx1 \div 10 $
nm$^{-1}$) that is particularly interesting as it is the region
where the transition from a genuine collective behaviour of the
density fluctuation fades in the single particle dynamics. Such
kinematics limitations have been recently overcome by the
Inelastic X-ray Scattering (IXS) technique. Using this techniques
it has been possible to firmly establish few common features of
the collective dynamics in glasses in the mesoscopic region
\cite{Science}. In particular, beside from specific {\it
quantitative} difference among different systems, all the
investigated glasses show some {\it qualitative} common features
that can be summarized as follows: {\it i)} there exist
propagating acoustic-like excitations up to a max $Q$ value
($Q_m$) given by $Q_m \; a \approx 1 \div 3$ showing up as more or
less defined Brillouin peaks at $\Omega(Q)$ in the $S(Q,\omega)$
(here $a$ is the average inter-particle distance). The specific
value of $Q_m\;a$ result to be correlated with the fragility of
the glass ; {\it ii)} the slope of the (almost) linear
$\Omega(Q)$ vs $Q$ dispersion relation in the $Q\rightarrow 0$
limit extrapolates to the macroscopic sound velocity; {\it iii)}
the broadening of the Brillouin peaks, $\Gamma(Q)$, follows a
power law, $\Gamma(Q)=D Q^\alpha$, with $\alpha\approx 2$ within
the statistical uncertainties; {\it iv)} the value of $D$ does not
depend significantly on temperature, indicating that the
broadening (i.e. the sound attenuation) in the high frequency
region does not have a dynamic origin, but rather that it is due
to the topological disorder \cite{glylowT}.

From a theoretical point of view, two main theories have been
developed for the study of supercooled and glassy systems, one
based on first principle computations of the equilibrium
thermodynamics~\cite{mezpar} and a second one, the Mode Coupling
Theory (MCT)~\cite{mct1,mct2}, dynamical in nature. Very recently
they have been complemented with the interpretation of the
thermodynamics and of the slow structural dynamics in terms of
the topological properties of the underlying Potential Energy
Surface (PES)~\cite{pes0,pes1,pes3}. Such
interpretation is based on the concept of {\it inherent
structures}, introduced many years ago by Stillinger and Weber
~\cite{pes2}: upon cooling, the system populates basins of the PES
whose (local) minima have depth increasing on lowering the
temperature. Moreover, in this framework, the dynamics of the
systems has been thought to be decomposed in a ``fast'' -high
frequency- vibrational dynamics describing the exploration of a
specific basin and a ``slow'' -diffusive- one, associated to the
exploration of different basins. In these studies a key role has
been played by computer simulations that, considering the
nowadays technology, permit a significant sampling of the PES
also at low temperature and, in principle, the study of the
dynamics of the model systems at every time scale.

Among other, one not yet fully explained issue in the high
frequency collective dynamics of supercooled liquids and glasses
is the recent findings that the temperature dependence of the
excitations frequency at fixed $Q$ value shows a slope
discontinuity at a temperature, $T_x$, that is {\it larger} than
the calorimetric glass transition temperature $T_g$
\cite{cbreak,cinf}. This observation seems to connect to each
other the shape of the PES basins (which determines the
vibrational frequency and hence also the excitations frequency at
a given $Q$'s) with the depth of the minima explored at that
specific temperature. It is our aim to further investigate this
issue.

In this paper we show results, obtained by means of molecular
dynamics simulations, concerning the high frequency dynamics of a
realistic {\it flexible} model~\cite{otp_self,otp_coll,mythesis}
for the fragile glass former orthoterphenyl (OTP). Our aim is
then twofold. First we want to study the capability of the
utilized molecular model to reproduce the high frequency dynamics
of OTP by comparing the calculated dynamics structure factor with
the analogous experimental quantity. Once the model has been
validate, our aim is i) to extend the experimental data to other
$Q$'s and other thermodynamic points and ii) to provide an
interpretation of the experimental finding of the existence of a
slope discontinuity in the temperature dependence of the sound
velocity ~\cite{cbreak,cinf}.

The paper is organized as follows: in Sec.~\ref{computation} we
briefly describe the model and we define the dynamic structure
factor $S(Q,\omega)$ that will be compared with the experimental
results. In Sec.~\ref{results} we recall some of the theoretical
model used in the analysis of the dynamic structure factor in
liquids; we  compare our results with the experimental sets of
data~\cite{cbreak,cinf} in a large temperature and momentum range
finding a very good agreement; we discuss the connection between
our findings and the PES approach to the dynamics. Finally in
Sec.~\ref{conclusions} we discuss the results obtained and we draw
some conclusions.

\section{COMPUTATIONAL DETAILS}
\label{computation}

In our model \cite{otp_self}  the OTP molecule is constituted by
three rigid hexagons (phenyl rings) of side $L_{a}=0.139$ nm. Two
adjacent vertices of the {\it parent} (central) ring are bonded
to one vertex of the two {\it side} rings by bonds whose length,
at equilibrium, is $L_{b}=0.15$ nm. In this scheme, each vertex of
the hexagons is thought to be occupied by a fictious atom of mass
$M_{CH}=13$ a.m.u. representing a carbon-hydrogen pair (C-H). In
the isolated molecule, at equilibrium, the two lateral rings lie
in planes that form an angle of about 54$^o$ with respect to the
central ring's plane.  The three rings of a given molecule
interact among themselves by an {\it intra-molecular} interaction
potential; such potential is chosen in such a way to give the
correct relative equilibrium positions for the three phenyl
rings, to preserve the molecule from "dissociation", and to
represent at best the isolated molecule vibrational spectrum. The
interaction among the rings pertaining to different molecules is
described by a site-site pairwise additive Lennard-Jones 6-12
potential, each site corresponding to one of the six hexagons
vertices. The details of the intra-molecular and inter-molecular
interaction potentials, together with the values of the involved
constants, are reported in Ref.~\cite{otp_self}. Previous studies
of the temperature dependence of the self diffusion
coefficient~\cite{otp_self} and of the structural ($\alpha$)
relaxation times~\cite{otp_self,otp_coll} indicate that the
introduced model is capable to quantitative reproduce the
dynamical behavior of the real system, but the actual simulated
thermodynamic temperature has to be shifted by $\approx$20 K
upward. In the following, as is our aim to compare the simulation
results with the experiments, the reported MD temperatures are
always shifted by such an amount.

We have studied a microcanonical (constant energy) system composed
by 108 molecules (324 rings, 1944 Lennard Jones interaction sites)
enclosed in a constant volume cubic box with periodic boundary
conditions. To integrate the equations of motion we have treated
each ring as a separate rigid body, identified by the position of
its center of mass $\bar R_{i}$ and by its orientation expressed
in terms of quaternions $\bar q_{i}$ \cite{allen}. The standard
Verlet leap-frog algorithm \cite{allen} has been used to
integrate the translational motion while, for the orientational
part, a refined algorithm has been used \cite{ruosam}. The
integration time-step is $\delta t$=2 fs, which gives rise to an
overall energy conservation better than 0.01 \% of the kinetic
energy.

We studied a wide temperature range spanning the liquid phase and
reaching the glass region. In the different temperature runs the
size of the cubic box has been chosen in order to keep the system
at the coexistence curve experimental density
\cite{density,giuphd}. At each temperature, after an
equilibration run (15 ns long) we start the calculation of the
molecular dynamics trajectory. It is worth to note that the
system in the low temperature side of the supercooled region and
in the glassy phase in not equilibrated, as the diverging
structural relaxation times do not allow simulation long enough
to reach the equilibrium condition. Nevertheless, as we are
interested to the high frequency properties, i.e. to those
properties deriving from the intra-basin dynamics of the system,
we think that it is not really important that the inter-basin
dynamics was really equilibrated. Actually, also in the real
experiments, whose outcome we are going to compare with the
simulation, one investigates the glassy phase in similar
non-equilibrium situation: the relaxation times in the glassy
phase are very long, even in comparison with the experimental
measuring times. During the MD evolution, lasting for $t_M$=640
ps, the configurations of the system are stored for subsequent
analysis every 20 (50) time step for the runs at $T>$ ($<$) 280
K. All the calculations have been performed on a cluster of four
$\alpha$-CPU with a frequency of 500 MHz; every nanosecond of
simulated dynamics needed approximately 24 hours of CPU-time.

From the stored configuration we calculate the dynamic structure
factor as measured in an Inelastic X-ray Scattering (IXS)
experiment~\cite{cbreak,cinf}. The IXS technique measures the
electron charge density fluctuation correlation function. As we
consider the phenyl ring as a rigid body -i.~e. we are not
interested in the intra-ring vibrational dynamics that takes
place at frequencies much higher than those investigated here- we
can consider as fixed the phenyl ring charge distribution. Within
this approximation, the scattering center can be considered as
the ring center of mass, and the effect of the spatial
distribution of the electron charge is summarized in a $Q$
dependent phenyl ring form factor not affecting the frequency
shape of $S(Q,\omega)$. Moreover, as we are not comparing neither
the calculated and measured absolute scattering intensities nor
their $Q$-dependence, we neglect hereafter the presence of such
a  form factor. Therefore, the appropriate dynamic structure
factor~\cite{hansen} to be compared with the IXS experimental
results is given by the power spectrum of the phenyl ring center
number density fluctuation $\delta\rho_{\bar Q}(t)$:
\begin{equation}
    \delta\rho_{\bar Q}(t)=
    \frac{1}{\sqrt{N}} \sum_{j}
    e^{i \bar Q \cdot \bar R_j(t)}
\label{rho}
\end{equation}
\begin{equation}
S(Q,\omega)=\frac{1}{t_{M}} \left \vert \int_0^{t_{M}} dt \; \;
       \delta\rho_{\bar Q}(t) \; \; e^{i \omega t} \right \vert^2.
\label{SQE}
\end{equation}
In this equation $t_M$ is the observation time for the variable
$\delta\rho_{\bar Q}(t)$. The dynamic structure factors have been
evaluated at the $Q$ values allowed by periodic boundary
condition of the simulation box: $\bar Q=2\pi/L \; (n,m,l)$, with
$n,m,l$ integers and has been sampled at circular frequency
between 0 and 22 ps$^{-1}$ with a step of $\approx$0.075
ps$^{-1}$. The smallest accessible $Q$ value is $\approx$1.8
nm$^{-1}$. 

\begin{figure}
\hbox to\hsize{\epsfxsize=1.0\hsize\hfil\epsfbox{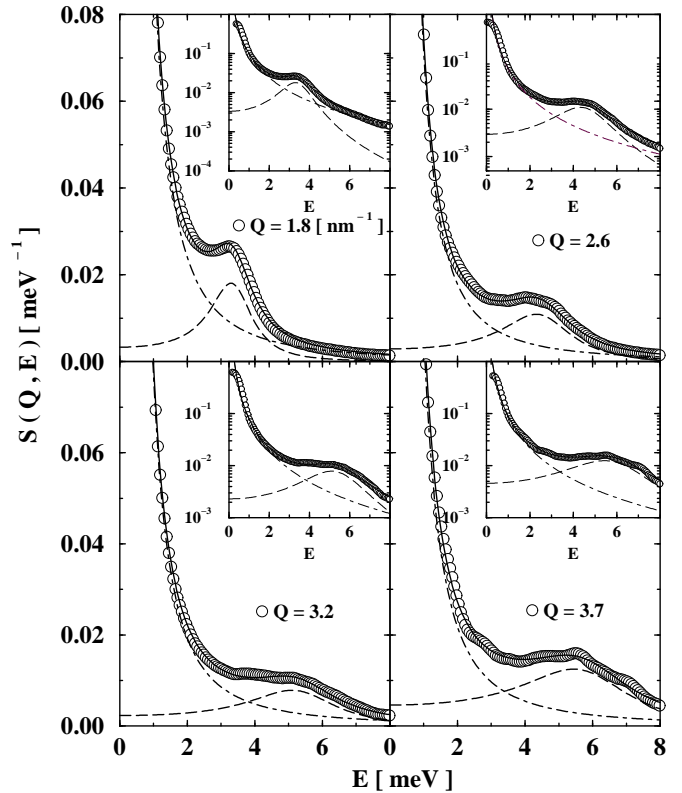}\hfil}
\vspace{.4cm} \caption {Selected examples of the $Q$ dependence of
the $S(Q,\omega)$ calculated from the MD runs (open circles) at
$T=$50 K and at the indicated $Q$ values. The data are shown in
linear and logarithmic scale, in the main figures and in the insets
respectively. The full lines represent the fit of the data to
Eq.~(\ref{DHO}), while the dashed (dot-dashed) lines is the
individual elastic (inelastic) contributions to Eq.~(\ref{DHO}).}
\label{fig1}
\end{figure}

To increase the statistics, the $S(Q,\omega)$ at
different $Q$ values have been binned in channels 0.2 nm$^{-1}$
wide, a value comparable to the experimental $Q$ resolution. The
frequency resolution, dictated by the time extension of the runs,
is $\Delta \omega$=0.15 (0.75) ps$^{-1}$ for $T>$ ($<$) 280 K. The
choice of different frequency resolutions below and above 280 K
is due to the fact that below this temperature the width of the
central line (proportional to the inverse of the structural
relaxation time $\tau_\alpha$) becomes so small that is no longer
measurable. It is therefore preferable to relax the energy
resolution in order to increase the statistics of the data. Above
280 K, the width of the central line is measurable, but a ``good''
energy resolution is needed; this procedure produces, obviously,
data with much poorer statistics.

\section{RESULTS}
\label{results}

Selected $S(Q,\omega)$~\cite{nota_sqw} at $T$=50 K and at Q=1.8,
2.6, 3.2 and 3.7  nm$^{-1}$ are reported in Fig.~\ref{fig1} (open
circles). 

\begin{figure}
\hbox to\hsize{\epsfxsize=1.0\hsize\hfil\epsfbox{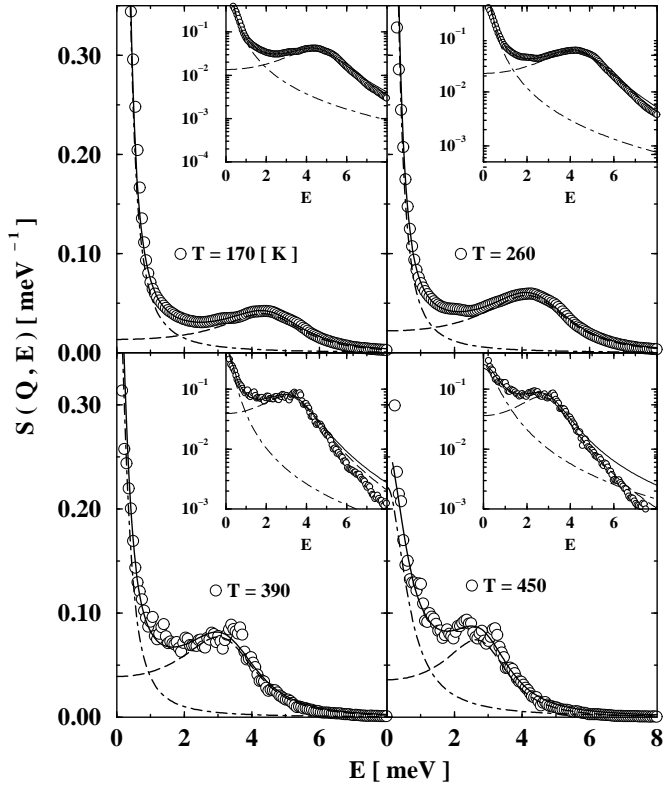}\hfil}
\vspace{.4cm} \caption {Selected examples of the $T$ dependence of
the $S(Q,\omega)$ at the fixed $Q$ value of 2.6 nm$^{-1}$.
Similarly to Fig.~\ref{fig1} the data are shown in linear and
logarithmic scale in the main figures and in the insets
respectively. The full lines represent the fit of the data to
Eq.~(\ref{DHO}), while the dashed (dot-dashed) lines is the
individual elastic (inelastic) contributions to Eq.~(\ref{DHO}).}
\label{fig2}
\end{figure}

\noindent In all the figures, to be consistent with the
experimental data, the $S(Q,\omega)$ have been reported as a
function of the energy ($E$) measured in meV rather than as a
function of the circular frequency $\omega$. For completeness we
recall that $\omega$[ps$^{-1}$]$\approx$1.61 $E$[mev]. As the
inelastic features appear as weak shoulder of an intense central
peak, for each $Q$ values the $S(Q,\omega)$ are shown both in
linear (main figures) and in logarithmic (insets) intensity
scale. We are interested in extracting the relevant parameters
from the calculated $S(Q,\omega)$. 

A formally exact way to treat
these functions goes through the description of the density
fluctuation correlation function $F(Q,t)=\langle \delta\rho_{\bar
Q}(t)\delta\rho^*_{\bar Q}(0)\rangle$, i.~e. the frequency
Fourier transform of $S(Q,\omega)$ in term of a generalized
Langevin equation \cite{baluc}:
\begin{equation}
    \label{lang}
    \ddot F(Q,t)+\omega^2_o F(Q,t)+
    \int_o^t m(Q,t-t') \dot F(Q,t') dt =0
\end{equation}
where $\omega^2_o$=$k_BTQ^2/MS(Q)$, $m(Q,t)$ is the "memory
function" and $S(Q)$ is the static structure factor.
This equation is ``exact'', but all the difficulties associated
to the calculation of the $S(Q,\omega)$ have been transferred to
the determination of $m(Q,t)$. The advantage of this formulation
stays in the fact that -independently from the choice of
$m(Q,t)$- the first two sum rules for $S(Q,\omega)$ are
automatically satisfied:
\begin{eqnarray}
    \int d\omega \; S(Q,\omega) &=& F(Q,t=0)= S(Q) \\
    \int d\omega \; \omega^2 \; S(Q,\omega) &=&
    -\ddot F(Q, t=0)= \frac{K_BT}{M}Q^2
    \nonumber
\end{eqnarray}
By Fourier transform of Eq.~(\ref{lang}), it is easy to show that:
\begin{eqnarray}
    S(Q,\omega)=\frac{\pi^{-1} \;
    S(Q) \; \omega_o^2 \; m^{\prime }(Q,\omega)}
    {\left[ \omega^2-\omega _o^2+
    \omega m^{\prime \prime }(Q,\omega)\right]^2
    +\left[ \omega m^{\prime }(Q,\omega)\right] ^2}
\label{sqwgenerale}
\end{eqnarray}
where $m^{\prime }(Q,\omega)$ and $m^{\prime\prime}(Q,\omega)$
are the real and imaginary part of the time Fourier transform of
the memory function. In the $\omega \tau_\alpha >>$1 limit, a
limit that is valid in all the investigated temperature
range~\cite{giuphd}, the memory function can be considered as the
sum of two contribution: a constant -which reflects the
$\alpha$-process frozen on the time scale of the sound waves- plus
a very fast decay at short time. The latter contribution to the
memory function -often referred to as ``microscopic'' or
``instantaneous''- is usually represented as a delta-function.
Introducing the two constants representing the area of the
``instantaneous'' process and the long time limit of the memory
function, $2\Gamma(Q)$ and $\Delta_\alpha^2(Q)$, the memory function is
approximated by
\begin{equation}
m(Q,t)=2\Gamma(Q)\delta(t)+\Delta_\alpha^2(Q)
\label{memDHO}
\end{equation}
and therefore, using Eq.~(\ref{sqwgenerale}), the $S(Q,\omega)$ reduces to:
\begin{eqnarray}
&&S(Q,\omega)=S(Q) \times\nonumber\\
&&\left [ f_Q \delta(\omega) +
(1-f_Q) \frac{1}{\pi}
       \frac{\Omega^2(Q)\Gamma(Q)}
       {(\omega^2-\Omega^2(Q))^2+\omega^2\Gamma^2(Q)}
       \right ]
\label{DHO}
\end{eqnarray}
where $\Omega(Q)=\sqrt{\Delta_\alpha^2(Q)-\omega_o^2}$ and
$f_Q=1-\omega_o^2/\Omega^2(Q)$. This expression is the sum of an
elastic line (the frozen $\alpha$ process) and of an inelastic
feature which is formally identical to a Damped Harmonic
Oscillator (DHO) function; the central line accounts for a
fraction $f_Q$ -the Debye-waller or non-ergodicity factor- of the
total intensity.

Recently it has been shown both via MD \cite{relax_harm} and via IXS
\cite{litio} that the ``microscopic'' contribution to the memory
function cannot be represented by a delta-function. There are, indeed,
clear indications that this "microscopic" part is responsible for
a positive dispersion of the sound velocity in "harmonic" model glasses
\cite{relax_harm}, and is responsible for the majority of the positive
dispersion of the sound velocity in liquid lithium \cite{litio}.
The simplest generalization of the delta function appearing in
Eq.~(\ref{memDHO}) considers a simple Debye-like "microscopic''
contribution to the memory function:
\begin{equation}
m(Q,t)=\Delta_\mu^2(Q) e^{-t/\tau_\mu}+\Delta_\alpha^2(Q).
\label{memDEB}
\end{equation}
\begin{figure}
\hbox to\hsize{\epsfxsize=1.0\hsize\hfil\epsfbox{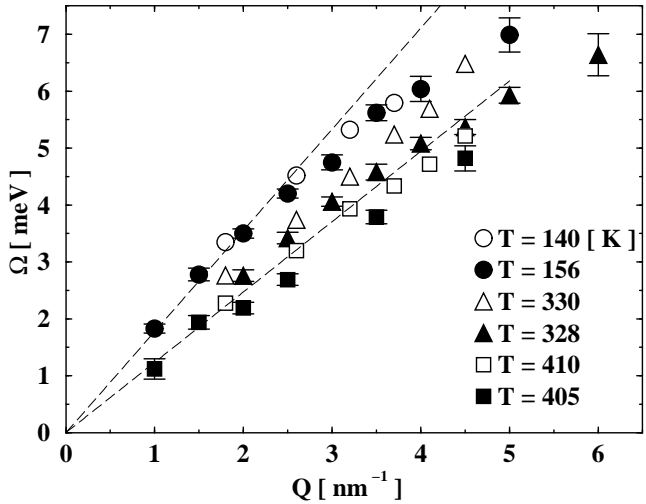}\hfil}
\vspace{.4cm}
\caption {$Q$ dependence of the excitations frequency $\Omega(Q)$
at the indicated selected temperatures (open symbols for MD data and
closed symbols for IXS data) as derived from the fit of
the $S(Q,\omega)$ to Eq.~(\ref{DHO}). The dashed lines are guide to the eye.
}
\label{fig3}
\end{figure}
\begin{figure}
\hbox to\hsize{\epsfxsize=1.0\hsize\hfil\epsfbox{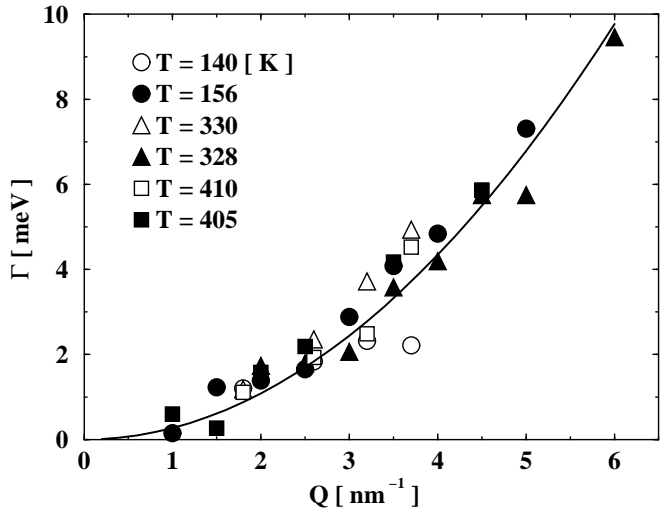}\hfil}
\vspace{.8cm} \caption {$Q$ dependence of the excitations
broadening $\Gamma(Q)$ at the indicated selected temperatures
(open symbols for MD data and closed symbols for IXS data) as
derived from the fit of the $S(Q,\omega)$ to Eq.~(\ref{DHO}). The
full line is a quadratic ($\Gamma(Q)=dQ^2$) fit to the whole set
of data. } \label{fig4}
\end{figure}
\noindent The quantity $\tau_\mu$ entering in this equation is the
characteristic time of the ``microscopic''  contribution to the
memory function. Obviously, whenever $\Omega \tau_\mu <<1$,
Eq.~(\ref{memDEB}) produces for $S(Q,\omega)$ a DHO-like
lineshape with $\Omega(Q)=\sqrt{\Delta_\alpha^2(Q)+\omega^2_o}$
and $\Gamma(Q)=\Delta^2_\mu(Q) \tau_\mu$. 

In analyzing the IXS
spectra of OTP, the author of Ref.~\cite{cbreak} used the
simplified DHO expression for the $S(Q,\omega)$ without noticing
any systematic deviation of the experimental data from the
fitting lineshape.
\begin{figure}
\hbox to\hsize{\epsfxsize=1.0\hsize\hfil\epsfbox{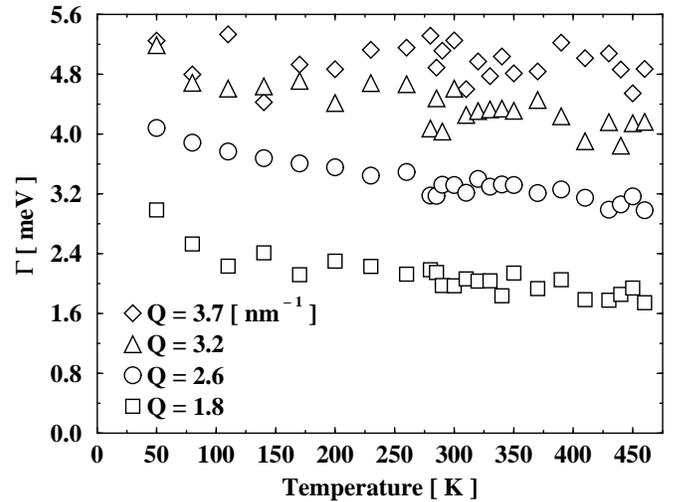}\hfil}
\vspace{.4cm}
\caption {$T$ dependence of the excitations broadening $\Gamma(Q)$
at the indicated selected $Q$ values as derived from the fit of
the $S(Q,\omega)$ to Eq.~(\ref{DHO}).
}
\label{fig5}
\end{figure}
\noindent Here, in order to better compare the MD results with the IXS one,
we will follow the same procedure and analyze the calculated
$S(Q,\omega)$ to the lineshape reported in Eq.~(\ref{DHO}). It is
worth to note, however, that the MD data, less noisy and less
affected by the resolution function than the experimental data,
actually allows to establish that the $S(Q,\omega)$ obtained by
the use of the memory function in Eq.~(\ref{memDEB}) produces a
better fit of the data. The fits to the MD $S(Q,\omega)$ using
Eq.~(\ref{DHO}) has been made via the minimization of the
standard $\chi^2$ function, and the results are reported in
Fig.~\ref{fig1} as full lines. The individual contributions
(elastic and  inelastic) are also shown (dashed and dot-dashed
lines). The agreement between the MD data and the fitting function
is satisfactory for all the investigated $T$ and for all the
investigated $Q$. An example of the $T$ dependence of the
$S(Q,\omega)$ at $Q=2.6$ nm$^{-1}$ is reported in Fig.~\ref{fig2},
again in linear (main figure) and logarithmic (inset) scale and
together with the best fit to a DHO lineshape.

The $Q$-dependences of the DHO parameters $\Omega(Q)$ and
$\Gamma(Q)$ together with the experimental results of
Refs.~\cite{cbreak,cinf} at selected temperatures are reported in
Figs.~\ref{fig3} and~\ref{fig4} respectively~\cite{nota_gamma}.
These figures confirms the experimental findings (see Fig.~1
in~\cite{cbreak} and Fig.~2  in~\cite{cinf}) that: i) The
excitations frequencies $\Omega(Q)$ show a clear $Q$ dependence,
which approach a linear behavior at  small $Q$'s and bent down at
increasing $Q$ values; ii) The slope at small $Q$ of $\Omega(Q)$
shows a marked $T$ dependence together with its overall shape;
iii) The parameters $\Gamma(Q)$ follow a $Q^2$ behavior; and iv)
The temperature dependence of $\Gamma(Q)$ is much less pronounced
than that of $\Omega(Q)$.

The latter $T$ dependence is better emphasized in Fig.~\ref{fig5},
where the parameter $\Gamma(Q)$ is reported -for the indicated
selected $Q$ values- at all the investigated temperatures.
\begin{figure}
\hbox to\hsize{\epsfxsize=1.0\hsize\hfil\epsfbox{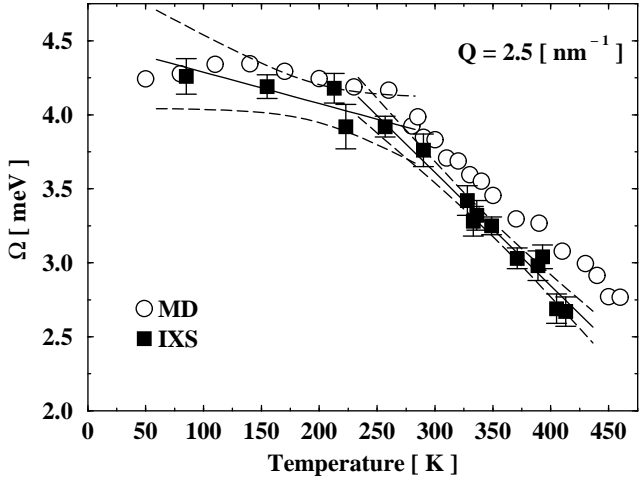}\hfil}
\vspace{.4cm} \caption { $T$ dependence of the excitations
frequency $\Omega(Q)$ at $Q$=2.6 nm$^{-1}$ (scaled at $Q$=2.5
nm$^{-1}$ as explained in~\protect\cite{nota_q}) and as derived
from the fit of the calculated $S(Q,\omega)$ to Eq.~(\ref{DHO})
(open circles). Also reported are the values of $\Omega(Q)$ at
$Q$=2.5 nm$^{-1}$ determined from the IXS experiment (full
squares) \protect\cite{cbreak}. The full lines are the best fit
to the experimental points in the low- and high-temperature
regions: they cross at the temperature $T_x$. The dashed lines
indicate the limit of $\pm\sigma$ predictions band.}
\label{fig6}
\end{figure}
\noindent A decrease (line narrowing) of $\Gamma(Q)$ of about 20\% is
observed between the glassy phase at 50 K and the liquid phase
(450 K). This decreasing behavior can be explained remembering
that, in the $\Omega(Q) \tau_\mu << 1$ limit, the parameter
$\Gamma(Q)$ is given by:
\begin{equation}
\Gamma(Q)=\Delta_\mu^2(Q)
\tau_\mu=(\Omega_\infty^2(Q)-\Omega^2(Q))\tau_\mu
\label{gamma}
\end{equation}
and considering that both $\Omega(Q)$ and $\Omega_\infty(Q)$ are
linear in $Q$ and proportional to the sound velocities
appropriate to frequencies always much larger than
$1/\tau_\alpha$ but respectively smaller and larger than
$1/\tau_\mu$.  The decrease of $\Gamma(Q)$ with increasing $T$
reflects therefore the decrease of the sound velocities with
increasing $T$. It is also important to point out that
Eq.~(\ref{gamma}) gives also an explanation for the observed
$Q^2$ behavior of $\Gamma(Q)$, when the reasonable hypotheses of
a weak $Q$ dependence of $\tau_\mu$ and of the sound velocities
are made. Overall the MD results reported in
Figs.~\ref{fig2}~-~\ref{fig5} are in agreement with the IXS
experimental findings and give further support to the picture
where propagating high frequency sound modes exist in glasses and
liquids. These excitations are characterized (in the small $Q$
limit) by a linear and a quadratic $Q$ dependence of the
excitations frequency and sound absorption coefficient
respectively.

To be more quantitative, we report in Fig.~\ref{fig6} the
comparison of the $T$ dependence of the $\Omega(Q)$ parameter at
fixed $Q$ value of 2.5 nm$^{-1}$ \cite{nota_q} as derived from
the DM and IXS data. The two set of data are in satisfactory
agreement.
\begin{figure}
\hbox to\hsize{\epsfxsize=1.0\hsize\hfil\epsfbox{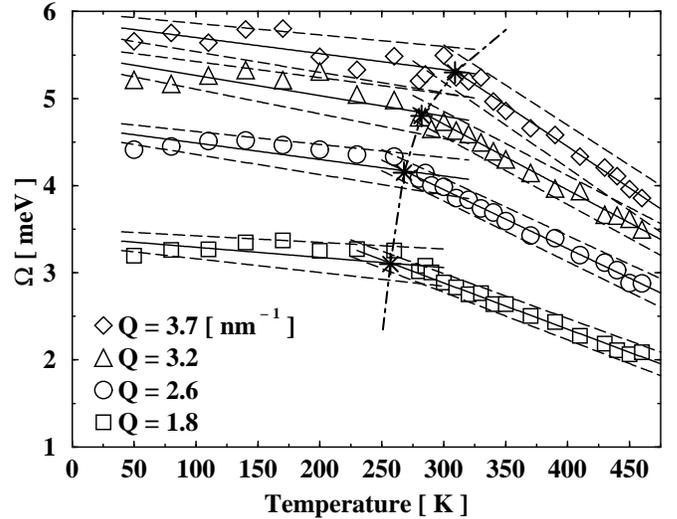}\hfil}
\vspace{.4cm} \caption {$T$ dependence of the excitations
frequencies $\Omega(Q)$ at the indicated $Q$
values derived from the fit of the calculated $S(Q,\omega)$ to
Eq.~(\ref{DHO}) (open symbols). The full lines are the best fit
to the points in the low -and high- temperature regions: they
cross at the temperature $T_x(Q)$, indicated by stars. The dashed
lines  indicate the limit of $\pm\sigma$ predictions band: their
lowest and highest $T$ crossing are used to define the error bar
for $T_x(Q)$. The dot-dashed line is a guide to the eye and
connects the different discontinuity temperature.} \label{fig7}
\end{figure}
\noindent The MD data lie slightly above the experimental ones, a feature
that we ascribe to the non perfect tuning of the molecular
potential model. It is important to underline that, similarly to
the experimental data,  also the MD derived $\Omega(Q=2.5\;
nm^{-1})$ show a discontinuity of slope at a temperature around
$T_g$. A discontinuity that also in the MD data appears to be
located at a temperature slightly above the calorimetric glass
transition temperature. However, the single set of data at
$Q=2.5$ nm$^{-1}$ is not conclusive. A more complete picture can
be derived from the data reported in Fig.~\ref{fig7}, where the
$T$ dependence of $\Omega(Q)$ is reported for four different $Q$
values. Here one can clearly see that the temperature $T_x(Q)$
where is observed the slope discontinuity shows a well defined
$Q$ dependence, and seems to approach $T_g$ decreasing $Q$. This
observation, obviously, indicates that the shape of the dispersion
relation $\Omega(Q)$ is modified by changing $T$. Therefore it is
not possible to scale all the $\Omega(Q)$ vs. $Q$ curves on top
of each others by a factor, and it is important ot study the
$Q$$\rightarrow$0 limit of these dispersion relations. This
extrapolation is illustrated in Fig.~\ref{fig8} where, at four
selected temperatures, the ``apparent'' sound velocity
$c(Q)=\Omega(Q)/Q$ is reported vs. $Q$. A fit of these apparent
velocities to a quadratic function, $c(Q)=c(Q=0)+a Q^2$ is then
performed (dashed lines in Fig.~8). The $Q=0$ sound velocity,
obtained by such an extrapolation are reported in Fig.~\ref{fig9},
together with the similar quantity derived from the IXS data
\cite{cinf}.
\begin{figure}
\hbox to\hsize{\epsfxsize=1.0\hsize\hfil\epsfbox{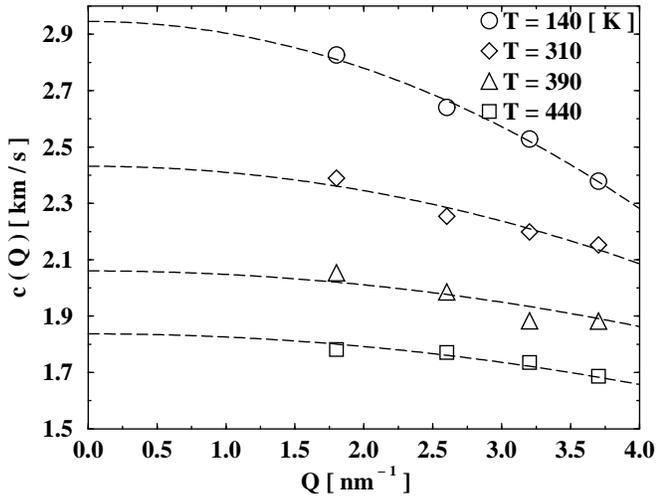}\hfil}
\vspace{.4cm}
\caption {The apparent sound velocities $c(Q)=\Omega(Q)/Q$
are reported as a function of $Q$ at the four indicated temperatures.
The dashed lines are the best fits of $c(Q)$ to a quadratic function,
$c(Q)=c(Q=0)+aQ^2$, and are used to determine the $Q$=0 limit of the
sound velocity.}
\label{fig8}
\end{figure}

\begin{figure}
\hbox to\hsize{\epsfxsize=1.0\hsize\hfil\epsfbox{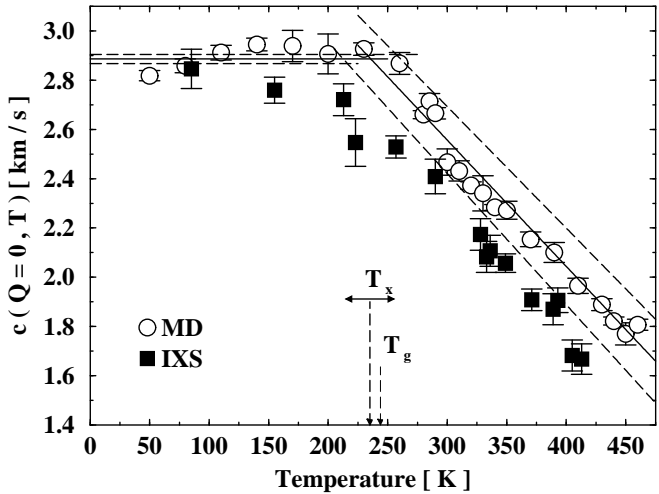}\hfil}
\vspace{.4cm}
\caption {$T$ dependence of the sound velocity at $Q$=0
derived from the fits as in Fig.~\ref{fig8} (open circles). Also reported are
the corresponding values  determined from the IXS
experiment (full squares) \protect\cite{cinf}. The full lines are the
best fit to the MD points in the low- and high-temperatue
regions: they cross at the temperature $T_x(Q=0)$. The dashed lines
indicate the limit of $\pm\sigma$ predictions band.
}
\label{fig9}
\end{figure}
\noindent Again we found a good agreement between simulated and experimental
data, and again the two sets of data show a slope discontinuity
at a comparable temperature, that now {\it coincide within the
statistic uncertainties} with $T_g$.

\begin{figure}
\hbox to\hsize{\epsfxsize=1.0\hsize\hfil\epsfbox{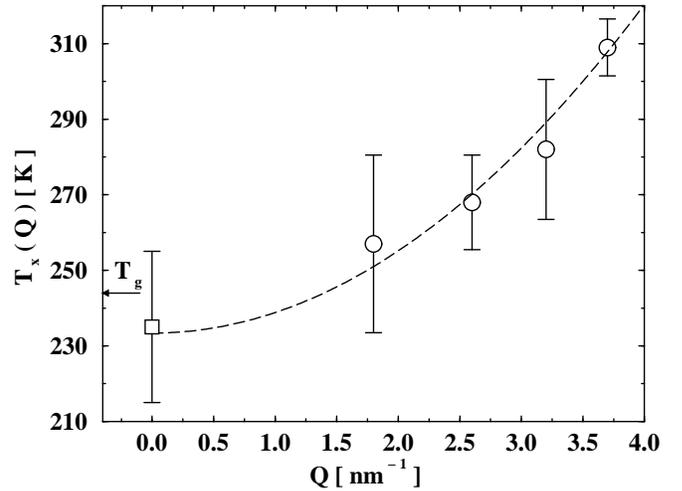}\hfil}
\vspace{.4cm}
\caption {$Q$ dependence of the discontinuity temperature $T_x(Q)$
derived from Fig.~\ref{fig7} (open circles) and from Fig.~\ref{fig9}
(open square). The error bars are determined by the crossing of the
$\pm\sigma$ prediction bands. The dashed line is a guide to eye. The
orizontal arrow indicate the calorimetric glass transition temperature.
}
\label{fig10}
\end{figure}

The overall picture is summarized in Fig.~\ref{fig10}, where the
discontinuity temperature $T_x(Q)$ is reported as a function of
$Q$. Despite the large statistical error bars, it is evident that
$T_x$ has a marked $Q$ dependence and, in particular, it is clear
that it extrapolates to a temperature compatible with $T_g$ as
$Q$ goes to zero.

This result can be interpreted in the framework of the current
understanding of the dynamics of supercooled liquids. Numerous MD
simulations study~\cite{pes0,pes1,pes3} have recently pointed out the
validity of the description of the supercooled liquid state in
terms of the formalism initially introduced by Stillinger and
Weber~\cite{pes2}: the point that represent the system in the
3$N$ configurational space moves, in this space, performing a
``fast'' vibrational dynamics around a given minima of the
potential energy hyper-surface and occasionally -with a rate
dictated by the structural relaxation time $\tau_\alpha$- perform
jumps between basins pertaining to different minima~\cite{pes0}. This
``naive'' description acquires more physical meaning with the
demonstration \cite{aging1,aging2,evst1,evst2} 
that the two dynamical regimes are
almost decoupled, to the extent that the aging processes in
glasses can be quantitatively described supposing that the two
subsystem are equilibrated at different temperatures.
In this framework, it has been demonstrated \cite{evst1,evst2}
that, approaching the glass transition temperature, the
representative point visits minima of lower and lower potential
energy, and exist a strict relation between the temperature and
the energy of the minima visited. As demonstrated in the case of
Binary Mixture Lennard Jones (BMLJ) \cite{evst1,evst2} systems,
different minima have also -on average- different metric
properties, as for example the curvatures at the minimum point
and therefore the distribution of vibrational eigenfrequencies.

Our findings on the $Q$-dependence of the discontinuity
temperature $T_x$ can be interpreted within this framework.
Similarly to Lennard-Jones systems, also in the case of OTP we
found that the modification of the eigenfrequencies (local
curvatures) does not follow a simple scaling, but there is a
deformation of the shape of the density of vibrational states in
changing the energy of the minima. In particular, upon cooling,
i.~e. on going to lower and lower energy minima, the highest
curvatures (highest frequency, highest $Q$) reach their limiting
values (those pertaining to the ``ideal'' glassy minima) before
than the lowest curvatures do.

\section{CONCLUSIONS}
\label{conclusions} In this paper we have reported results about
Molecular Dynamics calculation of the high frequency dynamics on
the flexible molecule model of orthoterphenyl introduced in
Ref.~\cite{otp_self}. The collective dynamics structure factor
has been determined in a wide exchanged momentum and temperature
range and has been compared to the its experimental measurements
performed by means of Inelastic X-ray Scattering
\cite{cbreak,cinf,giuphd}. The relevant parameters that describe
the overall spectral shape have been determined using the same
model already utilized in the analysis of the experimental data,
i.~e. the sum of an elastic contribution and a DHO lineshape. The
resulting parameters $\Omega(Q)$ and $\Gamma(Q)$ -the position
and broadening of the inelastic peaks respectively- are in good
agreement with the corresponding experimentally determined
quantities. Moreover the present simulation confirms the
existence in liquids and glasses of propagating sound waves at
high frequency and shows that: i) The excitation frequencies
$\Omega(Q)$ show a clear $Q$ dependence, which approach a linear
behaviour at  small $Q$'s and bent down at increasing $Q$ values;
ii) The slope at small $Q$ of $\Omega(Q)$ shows a marked $T$
dependence together with its overall shape; iii) The parameters
$\Gamma(Q)$ follows a $Q^2$ behaviour; and iv) the temperature
dependence of the high frequency sound attenuation coefficient,
related to $\Gamma(Q)$, is only weakly dependent on $T$. The
latter result further support the findings of a sound attenuation
mechanism not related to anharmonicity \cite{glylowT}.

More important, we find that -similarly to the experimental
findings of Ref.~\cite{cbreak}- the temperature dependence of the
excitation frequency at constant $Q$ value shows a slope
discontinuity at a temperature $T_x$ close to, but definitively
larger than, the calorimetric glass transition temperature. By
extending the $Q$ values where this analysis is performed, we
found clear evidence of the $Q$-dependence of $T_x$. In
particular, we find that $T_x(Q)$ approaches $T_g$ in the
$Q$$\rightarrow$0 limit. The existence of a $T$ dependence of the
$S(Q,\omega)$ peaks position can be explained by the results of
recent MD studies of Lennard-Jones systems where the
$T$-dependence of the curvatures of the PES at the inherent
structures were clearly evidentiated. The findings of a
$Q$-dependent transition temperature is an indication of the fact
that the density of vibrational states does not change according
to a frequency scaling law. Rather it is deformed in such a way
that the highest frequency reach the limiting value pertaining to
the ``ideal'' glassy minimum before than the lowest frequency.
Finally, our results indicate a possible {\it experimental} way to
validate or disprove the PES-based interpretation of the glass
transition phenomenology.

\begin{center}
\bf ACKNOWLEDGEMENTS
\end{center}
We thanks R.~Di~Leonardo and T.~Scopigno for intensive discussions during
the data analysis.

\end{multicols}

\begin{references}
\bibitem{nota_sqw} It is worth noting that the following general
discussion of the models for the dynamic structure factor will be
given in terms of the frequency $\omega$, at variance with all
the available experimental data which express the energy in meV.
In view of the comparison between MD and the experimental results
all the data points in the pictures are expressed in meV.
\bibitem{nota_gamma} It is worth noting that
sometimes in the literature in Eq.~(\ref{memDHO}) appears the
term $\Gamma$ instead of $2\Gamma$; this is actually the case of
Refs.~[3,4]. In order to preserve the consistency with
Eq.~(\ref{memDHO}) we then show both the actual experimental and
MD values for $\Gamma$ multiplied by 2.
\bibitem{nota_q}
The experimental data were collected at $Q$=2.5 nm$^{-1}$, while
the present MD data are calculated at  $Q$=2.6 nm$^{-1}$. To
account for the small differences in the $Q$ values -and
therefore in the $\Omega$ values- we have scaled the MD
calculated $\Omega(Q)$ by the factor 2.5/2.6, appropriate for a
linear $Q$ dependencies of the excitation frequency.
\end{references}
\end{document}